\documentclass[preprint,authoryear]{elsarticle}

\usepackage{graphicx}
\usepackage{amssymb}

\usepackage{lineno}

\newcommand\aap{A\&A}                
\newcommand\apj{ApJ}                 
\newcommand\apjl{ApJ}                
\newcommand\nat{Nature}              
\newcommand\pasj{PASJ}               
\newcommand\solphys{Sol.~Phys.}      
\newcommand\ssr{Space Sci. Rev.}     

\newcommand{\zast}{    {\it Zeit. Fur. Astrophysik}} 
\newcommand{\phyt}{    {\it Physics Today }}

\def\apj{{\it Astrophys.~J.}}

\def\apjl{{\it Astrophys.~J.~Lett.}}

\def\nat{{\it Nature} }

\usepackage{subcaption}
\usepackage{pifont}
\usepackage{latexsym}
\usepackage{mathrsfs}
\usepackage{amssymb,amsbsy}
\usepackage{amsfonts}
\usepackage{graphicx}
\usepackage{tcolorbox}
\usepackage[colorinlistoftodos]{todonotes}
\usepackage{varwidth}
\usepackage{natbib}
\journal{Advances in Space Research}

\begin{document}

\begin{frontmatter}


\title{An analytical model of prominence dynamics}




\author[1]{Swati Routh\corref{cor1}}
\ead{swati.routh@jainuniversity.ac.in}
\author[2]{Snehanshu Saha}
\ead{snehanshusaha@pes.edu}
\author[1]{Atul Bhat}
\ead{mail@atulbhats.com}
\author[1]{Sundar M.N }
\ead{sundar.mn@gmail.com}

\cortext[cor1]{Corresponding author}
\address[1]{Department of Physics, Center for Post Graduate Studies, Jain University, No.18/3, 9th Main Road, Jayanagar, Bangalore, Karnataka 560011}
\address[2]{Department of Computer Science and Engineering, PES University South Campus, Hosur Road,  Bangalore, Karnataka 560100}

\begin{abstract}

Solar prominences are magnetic structures incarcerating cool and dense gas in an otherwise hot solar corona.  Prominences can be categorized as quiescent and active. Their origin and the presence of cool gas ($\sim 10^{4}$ K) within the hot ($\sim 10^{6} K$) solar corona remains poorly understood.  The structure and 
dynamics of solar prominences was investigated in a large number of observational 
and theoretical (both analytical and numerical) studies. In this paper, an analytic model of quiescent solar prominence is developed and used to demonstrate that the prominence velocity increases exponentially, which means that some gas falls 
downward towards the solar surface, and that Alfv\'en waves are naturally present 
in the solar prominences.  These theoretical predictions are consistent with the 
current observational data of solar quiescent prominences.

\end{abstract}


\end{frontmatter}


\section{Introduction}

It is a well known fact that solar prominences are cool, dense plasma clouds composed of small-scale ever-changing threads of fibrils, embedded in the hot solar corona \citep{Anderson1989}, \citep{Berger96}. The prominence plasma is in nearly equilibrium state supported by the magnetic field against gravity  \citep{Kippenhahn57,Raadu73}. 
\\
Quiescent prominences are large and appear as thin vertical sheets endowed with fine filamentary structure. These prominences display minor changes over a period of time (days)\citep{DavidW98}. Irrespective of the  "quiescent" phrase , these prominences display remarkable mass motion when observed in high resolution H$\alpha$ movies. These filaments possess the solar material concentrated as rope-like structures with diameter less than 300 km. 
\\
Primarily, two types of topology have been suggested for supporting prominences that are related to magnetic fields. The first one was put forward by Kipppenhahn and Schluter in 1957 \citep{Kippenhahn57}. Kuperus and Tandberg-Hanssen proposed the latter in 1967 \citep{Kuperus67}. This was developed further by Kuperus and Raadu (K-R) in 1973 \citep{Raadu73}. In the Kuperus-Schlitter (K-S) model, the prominence material sits on top of the field lines supported by the normal polarity field. The K-R model suggests that the prominence is embedded in an inverse polarity field. Simply stated, a prominence is considered as a sheet of plasma, erected in the corona, above a magnetic neutral line.
\\
Prominences are highly dynamical structures exhibiting flows in H$\alpha$, UV and EUV lines. The study of these flows improve our understanding of prominence formation and stability, the mass supply and the magnetic field structure of the prominence imparting great interest to these topics. A complex dynamics with vertical down flows, up flows and horizontal flows is observed in the H$\alpha$ lines and quiescent limb prominences (\citep{Chae08}, \citep{Engvold85}, \citep{Kubota and Uesugi86}, \citep{Lin03}, \citep{Zirker98}). The velocity of these flows lies between a range of 2 and 35 km/s, while in EUV lines, these flows seem to be of slightly higher velocity. The pertinent aspect of these observations correspond to various temperatures indicating the speed of flow corresponding to different parts of the prominence.  These flows seem to be field aligned due to the filament plasma.Vertical filamentary downflows often have been observed in vertically striated or 'hedgerow' prominences \citep{Engvold76}, \citep{Martres81} as well as vortices \citep{Liggett84}. Explaining these observations of vertical and rotational flows with existing theoretical MHD models is one of the major goals of prominence's investigations.

In more recent numerical studies performed by Terradas et. al. (\cite{Terradas}), the MHD equations have been solved and time evolution of solar quiescent prominences embedded in sheared magnetic arcades has been investigated.  Moreover,  \cite{Terradas01} have studied solar active prominences embedded in magnetic flux ropes. The authors have shown that prominences may originate in the solar photosphere and presented their evolution through the solar atmosphere. The physical properties of the solar prominences and the existence of oscillations associated with such prominences resulting from numerical simulations have also been presented and discussed. 

In this paper, we develop an analytical approach to investigate the dynamics of solar quiescent prominences by considering a simple model that is suitable for such an  analytical treatment. The main theoretical results obtained from the model are: 
\begin{itemize}
\item an exponential increase of the prominence velocity within very short time (few minutes) and then resuming the motion with a uniform velocity;
\item the downfall of cool gas and neutral material toward the solar surface, which
is consistent with the observational data;
\item the theoretical evidence for the existence of Alfv\'en waves responsible for driving oscillations observed in solar quiescent prominences
\end{itemize}

The paper is organized as follows: Section 2 presents the model of the prominence, 
based on the MHD equations; this is followed by the obtained results and conclusion 
in Sections 3 and 4 respectively.

\section{MHD equations}

\begin{figure}[h!]
 \centering
    \includegraphics[height=150px]{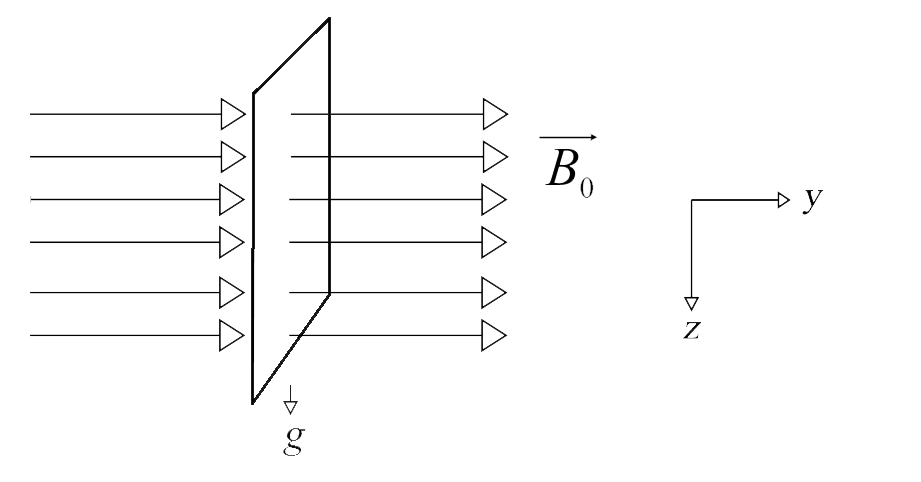}
\caption{Schematic representation of prominence plate along with conditions}
  \label{fig:prob}
\end{figure}

\par
H$\alpha$ photographs of quiescent solar prominences above
the solar limb often show evidence of the prominence plasma assuming the form of vertically oriented, narrow filaments (\cite{Tandberg-Hanssen}). Vector magnetic fields, using the Hanle effect, have been used  to observe the prominence plasmas. 
Such observations have helped establishing the fact that the magnetic fields inside 
the prominences are horizontal, binding across slab like macroscopic prominence while the principal field component remains parallel to the horizontal length of the prominence (\cite{Leroy}). The narrow vertical filaments are composed of pieces of plasma lined up vertically. H$\alpha$ observations also demonstrate that the filamentary structures of a quiescent prominence are not truly static \cite{Tandberg-Hanssen}, \cite{Zirin}. Here, we attempt to analyze the dynamics of such a prominence using the MHD equations. 

\par Suppose there exists an one dimensional infinite vertical rigid sheet of perfectly conducting massive material sitting in a perfectly conducting incompressible static fluid. Let the sheet be threaded by a uniform magnetic field that is perpendicular to the sheet. Gravity is assumed to be uniform and acts vertically to the prominence sheet. Gravity is neglected for the medium (assuming the medium's density to be significantly less compared to the sheet's density) but considered to be acting on the prominence thread. Let $\vec{B}(y,t) \hat{z}$ and $ \vec{v} = v(y,t) \hat{z}$ be the perturbations in the magnetic field and velocity of the medium respectively. The magnetic field acting on the prominence is described as
$\vec{B} = [ 0, B_0, \vec{B}(y,t)]$, where 
$\vec{B}(y,t) \hat{z} $ is perturbation in magnetic field, and 
$B_{0}$ is constant magnetic field in the $y$ direction, perpendicular to the sheet.  

The z-component of the MHD momentum equation for the medium outside the prominence sheet can be written in the following form,

\begin{equation} \rho_{0} \frac{{\partial v}}{\partial t}=  \frac{B_0}{4 \pi} \frac{\partial B}{\partial y}. \label{cpl1} \end{equation}
 and the z-component of MHD induction equation becomes,
\begin{equation} \frac{{\partial B}}{\partial t}=  B_0 \frac{\partial v}{\partial y} . \label{cpl2}\end{equation}
%

It must be pointed out that we have not used any small amplitude approximation to linearize the MHD equations in order to obtain the above equations.
\subsection{Alfven wave equations}
From Eqs. \ref{cpl1} \& \ref{cpl2}, the following equations are obtained,
\begin{equation}  \frac{\partial^2 v}{\partial t^2} = \frac{B_0^2}{4 \pi  \rho_0} \frac{\partial^2 v}{\partial y^2} \label{wave1} \end{equation}
\begin{equation}  \frac{\partial^2 B}{\partial t^2} = \frac{B_0^2}{4 \pi  \rho_0} \frac{\partial^2 B}{\partial y^2} . \label{wave2} \end{equation}

The above two equations (\ref{wave1} \& \ref{wave2}) are Alfven Wave equations, with Alfven wave velocity $V_A= \frac{B_0}{\sqrt{4 \pi \rho_0}}$.

\subsection{Solution to Wave Equations }
 \label{Solution to Wave Equations}
The following initial conditions are considered.
Initial Condition :  at $ t=0,  B=0$ and   $v=0$.    
The boundary Condition :  at $ y=0, v=u(t) \bigg|_{y=0} $, 
where $u(t)$ is the prominence sheet's vertical velocity component. 
\\
The general solution to the above wave equation (\ref{wave1}) is 
\begin{equation} B= F(y- V_A t)+G(y+ V_A t), \label{cond1} \end{equation}
The general solution of the one-dimensional wave equation is the sum of a right traveling function F and a left traveling function G. "Traveling function" implies that the shape of these functions (arbitrary) remains invariant with respect to y. However, the functions are translated left and right with time at the speed $V_A$ \citep{D'Alembert47}.
The solution, on substitution in eq. ~\ref{cpl1} \& ~\ref{cpl2}, yields the following: 
\begin{equation} v= \frac{1}{\sqrt{4 \pi \rho_0 }}[-F(y- V_A t)+G(y+ V_A t)]. \label{cond2} \end{equation}
\subsection{Momentum equation for prominence sheet}
It is imperative to understand the nature of the velocity in order to lend credence to the theory of the dynamics of the filaments.The momentum equation for the prominence sheet reads as:
\begin{equation} m_p \frac{\partial u}{\partial t} \bigg|_{y=0} =  m_p g - \int_{y=0-}^{y=0+} \frac{B_0 }{4 \pi} \frac{\partial B}{\partial y} \bigg|_{y=0}. \label{prom1} \end{equation}
where $m_p$ is the integrated mass density across the thickness of the thin sheet (mass density $\rho \times$  thickness $t$) and $u$ is the vertical component of prominence sheet's velocity .\\
The second boundary condition i.e normal component of the magnetic field being continuous across the plate, renders  eq. \ref{prom1} as
\begin{equation} m_p \frac{\partial u}{\partial t} \bigg|_{y=0} =  m_p g - \frac{B_0 B}{2 \pi} \bigg|_{y=0} . \label{prom2} \end{equation}
Incorporating the initial conditions in ~\ref{cond1} \& ~\ref{cond2}, 
 for t=0, we obtain $F(y)=G(y)$. The initial condition $ t=0, v=0 $ yields $ F(y)= - G(y) $.
Thus, 
\begin{eqnarray}
  \begin{array}{l l}
    G(y) =0 & \\
   F(y)=0 &
  \end{array}    
   t=0 , y>0  
\end{eqnarray}
Now consider, $(y-V_A t) = \xi $ and $(y+ V_A t) = \eta $. Similarly,
$\xi > 0$ implies $F(\xi) =0$ and \\
 $\eta > 0$ implies $G(\eta) = 0$.
Also note, for the 2 cases,
$ \xi > 0$ implies $y > V_A t$ which in turn, gives us  $F(\xi) =0$  and $ G(\xi) =0$.
This is a trivial solution.  
However, for $ \xi < 0$, we have $y<V_A t$ which implies $F(\xi) \neq 0$. This renders 
equation (\ref{prom2}) the following form: 


\begin{equation} \frac{m_p V_{A}}{\sqrt{4 \pi \rho_{0} }} \frac{d F(-V_A t)}{d t} =  m_p g - \frac{B_0 F(-V_A t)}{2 \pi} . 
\end{equation}

This is a differential equation of first order, which is solved by using integrating factor \citep{Saha11} as follows:
\begin{equation} I = exp \int (2 \rho_0 / m_p )dx. \label{firstorder} \end{equation} 
where $x=-V_{A}t$.
Applying the integrating factor yields the solution for the equation (\ref{firstorder}) as,
\begin{equation}\label{promvel} u = \frac{m_p g}{2 \rho_0 V_A} \bigg[1- exp (- \rho_0 V_A t / m_p)\bigg] \end{equation}
where, \\*
$u$ is the vertical component of prominence velocity,\\*  
$m_p$ is the product of prominence mass density and width of the thread ,\\*
$g$ is the acceleration due to gravity of the sun,\\*
$ \rho_0 $ 	is the density of the medium outside the filament,\\*
$ V_A $ is the Alfven wave velocity \\*
and $ t $ is the time.\\*

\section{Results}
The parameters used in our calculations have the following approximated values: the acceleration due to gravity on the solar surface $g$ is $27.42 \times 10^{3} cm s^{-2}$, the prominence filament density is $10^{11}$ particles/$cm^{3}$ and the coronal density is $10^{9}$ particles/$cm^{3}$ (\cite{Petrie}), and the filament width is $300 km$ (\cite{Low82}). The prominence velocities are determined for different Alfven wave velocities ranging from of $1 \times 10^{8} cm s^{-1}$ to $ 3 \times 10^{8} cm s^{-1}$. Plots for different cases are given below. 
 
\begin{figure}[h!]
 \centering
    \includegraphics[height=159px, width= 200px]{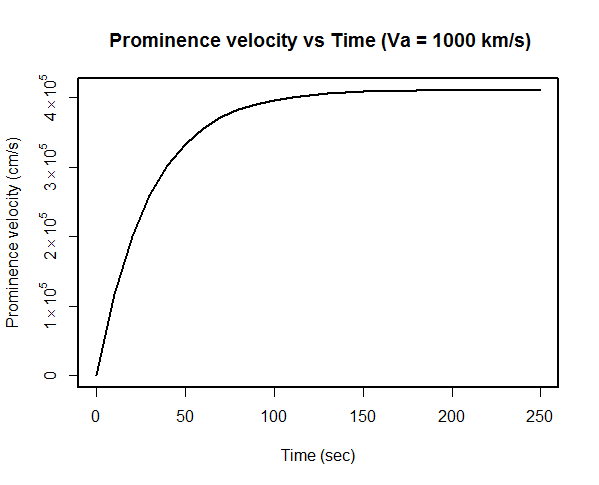}
    \includegraphics[height=159px, width= 200px]{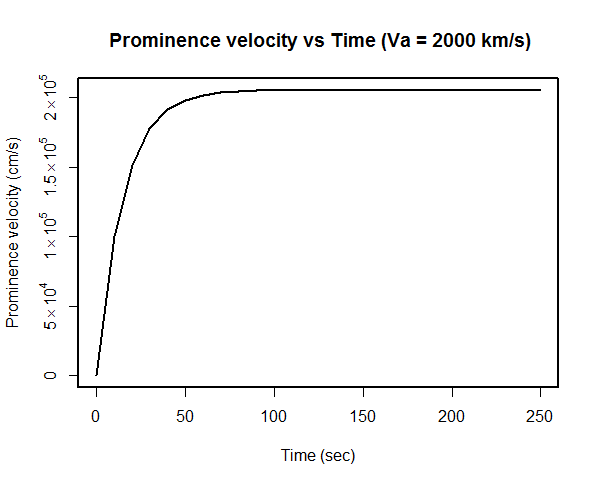}
    \includegraphics[height=150px, width= 200px]{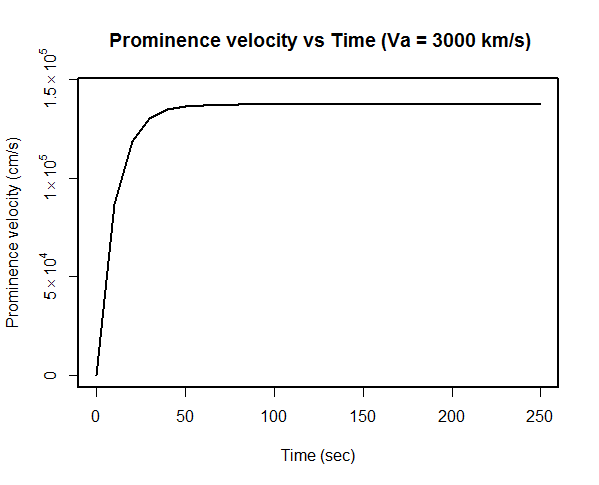}
\caption{Prominence velocity as a function of time (eq (\ref{promvel})) with Alfven wave velocities of $ 1000$ km $s^{-1}$,$ 2000$ km $s^{-1}$, $ 3000$ km $s^{-1}$ }
  \label{fig:promplot}
\end{figure}

The obtained results show that there is an exponential increase in the velocity of the prominence thread from the equilibrium state over a very short time (few minutes). Then, there is a downward fall (towards solar surface) of the prominence thread with 
a uniform velocity. Moreover, our results demonstrate the existence of Alfv\'en waves, which are likely to trigger oscillations commonly observed in solar prominences. The observational data collected by the Solar Optical Telescope on the Hinode satellite (e.g., \cite{Tsuneta}) are relevant to the theoretical results obtained in this paper because the velocity of prominences can be determined from these observations. The data also show the downward fall of a cool and neutral gas. The observed oscillations are either global or local (see \cite{Mackay} and references therein), and they are likely to be driven by MHD waves present within the prominences. The prominence velocity vs time is shown in fig.~\ref{fig:promplot}.

Fig.2 shows plots of the prominence velocity for different Alfven wave velocities. For $V_{A}=1000$ km/s, the uniform value of filament velocity is 4.11 km/s, whereas for $V_{A}=2000$ km/s it is 2.05 km/s and for $V_{A}=3000$ km/s, it is 1.37 km/s. This is in agreement with the observations.

The vertical down-flow of matter has been reported by Kubota and Uesugi (\cite{Kubota and Uesugi86}). They found that the downward motion is predominant in the observed stable filament and ranges up to 5.3 km/s. Engvold et. al. (\cite{Engvold76}) and references therein also reported the overall motion of prominence material is directed downward and measured flow velocities of 5-15 km/s. Berger et. al. \citep{Berger08} found down flows of bright knots less than 10 km $s^{−1}$ looking at Ca II H images in the line center. Using Hinode/SOT data Schmieder and colleagues (\cite{Schmieder10}) have shown that the horizontal velocity in the quiescent prominences can reach up to 11 km 
$s^{-1}$.

Our results also show Alfven waves (not necessarily small amplitude, as we did not apply linearized approximations) could be produced by the prominence filament's vertical motion through the uniform background magnetic field. These waves are mostly localized (as for $y > V_A t$ implies $F(\xi) =0$  and $ G(\xi) =0$, i.e. no waves can propagate beyond the distance $V_A t$  from the prominence axis). As already 
mentioned above, the waves drive global or local oscillations, whose presence in solar quiescent prominences has been confirmed observationally, (\cite{Mackay}) and references therein.

\section{Conclusion}
A simple MHD model of solar quiescent prominences was developed and used to determine the velocity of the prominences. The results obtained from this model were then compared to the available solar observations. The main theoretical prediction of our simple model is that the prominence sheets move at steady uniform downward velocities (few kms/s) within their planes, in agreement with the observations. An important feature of our simple model is the downfall of cool, dense and neutral gas (it must be neutral, so it can fall through approximately horizontal magnetic field lines) towards the solar surface. The falling gas may generate Alfv\'en waves, which could potentially drive global and local oscillations observed in solar prominences.  The recent observations (see Sections 1 and 3) give strong evidence for the existence of both the oscillations and waves.

\par Based on the above, we conclude that our simple analytical model of solar quiescent prominences describes at least some aspects of the dynamics of prominence, which is consistent with the available observed data. The model does have some predictive power, which is obviously limited because of the simplicity of our model. Nevertheless, the results of this paper may set up the baseline for future analytical work on solar prominences. In the near future, we intend to consider the localized bow like structure of magnetic field (\cite{Low82}) and try to determine the combined effect of these structures and the larger scale magnetic field on the production of oscillatory motions in solar prominences.

\section{Acknowledgement} We are extremely grateful to Dr. B.C.Low (HAO, Boulder) for his insightful suggestions. The communicating author wishes to acknowledge the support received from The Inter-University Center for Astronomy and Astrophysics (IUCAA), Pune, India during the time spent as visiting associate in the academic year 2016-17. The authors express sincere gratitude to the referees for meaningful suggestions that helped to improve the quality of the paper. We gratefully acknowledge the valuable suggestions by Dr. Zdzislaw Musielak (University of Texas at Arlington).
 
\section*{References}

\bibliographystyle{model1-num-names}
\bibliography{sample.bib}

\begin{thebibliography}{99}
\bibitem[\protect\citeauthoryear{D'Alembert}{1747}]{D'Alembert47} D'Alembert. 1747, Recherches sur la courbe que forme une corde tenduë mise en vibration (Researches on the curve that a tense cord forms [when] set into vibration), Histoire de l'académie royale des sciences et belles lettres de Berlin, 3, 214-219.
\end{thebibliography}



\end{document}